\documentclass[aps,prl,singlecolumn,showpacks,superscriptaddress]{revtex4-1}
\usepackage[dvipdfmx]{graphicx}  
\usepackage{dcolumn}   
\usepackage{bm}        
\usepackage{amssymb}   
\usepackage{braket}
\usepackage[dvipdfmx]{color}
\usepackage{soul}
\usepackage{amsmath,amssymb,bm}
\usepackage{color}

\begin{document}

\widetext
\setlength{\textwidth}{175.0mm}

\title{Nonlocal inelastic scattering of light: Enhanced and noiseless signals in remote-coupled optomechanical systems} 
\date{\today}

\author{Sho Tamaki}
\affiliation{Department of Materials Engineering Science, Osaka University, Toyonaka, Osaka 560-8531, Japan}
\altaffiliation{Current address: Niels Bohr Institute, University of Copenhagen, Copenhagen, Denmark}

\author{Tomohiro Yokoyama}
\email{tomohiro.yokoyama@mp.es.osaka-u.ac.jp}
\affiliation{Department of Materials Engineering Science, Osaka University, Toyonaka, Osaka 560-8531, Japan}

\author{Hajime Ishihara}
\affiliation{Department of Materials Engineering Science, Osaka University, Toyonaka, Osaka 560-8531, Japan}
\affiliation{Department of Physics and Electronics, Osaka Prefecture University, Sakai, Osaka 599-8531, Japan}
\affiliation{Center for Quantum Information and Quantum Biology, Osaka University, Osaka, Japan}

\begin{abstract}
The inelastic scatterings of matter systems, such as Raman scattering, contain rich information on mechanical vibrations like as
resonant frequencies, which lead to various applications, for example, a sensor for specific molecules.
However, observing output signals requires a sensitive setup because an inelastic signal is inherently weak and is disturbed by strong input. 
In this study, we theoretically investigate a physical scheme to avoid detrimental impact of the input by distancing it
from the emitter and greatly enhancing the output signals.
If two bodies are coupled mechanically and direct optical communication is forbidden, the nonlocal inelastic scattering signals can be considerably boosted.
We demonstrate this mechanism by considering coupled optomechanical systems as a typical example that enables control of the two-body interaction strength.
The results present a general scheme to boost nonlocal inelastic scattering for noiseless and pure signals.
\end{abstract}

\pacs{}
\maketitle

\section{Introduction}
\label{introduction}
Light-matter interaction causes inelastic scattering of light accompanied by the generation of mechanical vibration, namely Stokes and anti-Stokes scatterings.
The energy shift of photons owing to such scattering is an integer multiple of the mechanical vibration frequency.
The Stokes shift owing to the mechanical vibration of molecules provides rich information, e.g., chemical bonds, crystallinities, and concentrations of
solutions~\cite{doi:10.1146/annurev-physchem-032511-143807,doi:10.1146/annurev.anchem.1.031207.112814}.
Hence, it is an essential tool for spectroscopy.
However, there are bottlenecks for observing inelastic signals, namely, the signals are inherently weak; therefore, they require a physical mechanism to enhance them. 
In surface- and tip-enhanced Raman scattering microscopes, the electric field is strongly enhanced by the localized surface plasmon resonances
that are sustained near the metallic nanostructures, which enhances the inelastic scattering at a molecule~\cite{Nie1102,PhysRevLett.78.1667}.
These spectroscopies have successfully observed signals from single molecules using nanoscale gaps between the apexes of metallic structures. 
However, there are still existing issues, namely, a stronger electromagnetic field causes larger and broader noise, 
which might disturb sensitive detection, owing to the masking of the minute signals of inelastic scattering.

The purpose of this letter is to theoretically propose a nonlocal inelastic-scattering scheme
that generates an enhanced and noiseless signal; thus, it is free from the above-mentioned bottleneck.
This scheme spatially and energetically separates the signals of inelastic scattering from
an intensive input laser, which is necessary for strong light-matter interaction; however, it simultaneously generates large noise.
For the theoretical demonstration, we used an artificial quantum system, the optomechanical system (OMS)~\cite{PhysRevA.51.2537,RevModPhys.86.1391}, 
as an example of a system that exhibits Stokes and anti-Stokes scattering, where
the light-matter interaction is attributed to an optical force caused by the enhanced light in a cavity.
Several parameters of the OMS, for example, mechanical frequency, the strength of the light-matter interaction, and damping constants,
can be designed; therefore, the full merit of the proposed scheme is well demonstrated by this system.
OMSs enable the detection and monitoring of mechanical motion via the inelastic scattering of light,
which can be applied to higher resolution sensors and efficient information transport~\cite{Andrews:2014aa,Forsch:2020aa,Jiang:2020aa}.
However, stronger field incidence causes larger and broader noise,
which might disturb sensitive detection at the local OMS, owing to the masking of the minute signal of inelastic scattering.

\begin{figure}[t]
\centering
\includegraphics[width=1.0\linewidth]{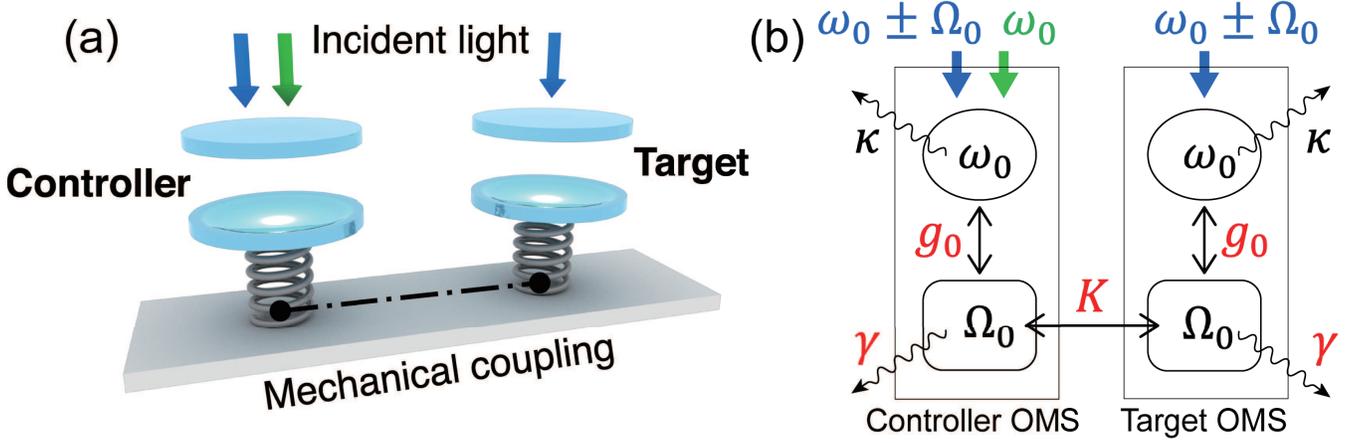}
\caption{
(a) Image of mechanically remote-coupled OMSs.
The c- and t-OMSs consist of one optical mode ($\hat{a}_\alpha$) and one mechanical vibration mode ($\hat{b}_\alpha$)
Here, $\alpha = {\rm c,t}$.
The phonons of the mechanical vibration modes in the OMSs are coupled to each other via a substrate, whereas the respective optical cavities are isolated.
Two-tone and monotone incident laser beams were applied to the c-and t-OMSs, respectively.
(b) Schematic summary of the parameters in the proposed system.
Colored parameters indicate the control parameters in this study.
}
\label{fig:model}
\end{figure}

To separate the output inelastic signals from the input, we consider remotely and mechanically coupled OMSs, for which an optical coupling between the OMSs is absent.
A mechanical coupling between OMSs has been experimentally examined to synchronize their mechanical vibrations~\cite{PhysRevLett.123.017402}
and achieve mechanical sensing and transduction~\cite{PhysRevApplied.14.014041}.
In theoretical studies, the mechanical coupling of OMSs has demonstrated a great potential to achieve
many-body interactions, which cause the phase transition~\cite{PhysRevLett.111.073603} and an emergence of topological nature~\cite{PhysRevX.5.031011}.
However, the nonlocal response through mechanical coupling has not been studied thus far.
In our system, one OMS is a controller under a two-tone laser irradiation, and the other is a target under a monotone laser. (see Fig.\ \ref{fig:model}(a).)
We observe the output signals at the target OMS (t-OMS), where nonlocal inelastic scattering is induced by
an interplay between the monotone input and phonons coming from the controller OMS (c-OMS).
This scheme separates the inelastic signal from noise and heat, and it might offer long coherence between the photon and phonon
because of the suppression of the heating noise due to two-photon absorption~\cite{6975027,Schneider:19}.
This is a critical advantage for various applications and the measurement of pure inelastic scattering.
Here, we wish to remark that the application of our proposed scheme is not limited to OMSs,
but it includes the general concept of generating nonlocal inelastic signals, thereby reducing detrimental impact of the strong input.

As depicted in Fig.\ \ref{fig:model}(b), the cavity of each OMS has a single optical mode described by $\hat{a}_\alpha , \hat{a}_\alpha^\dagger$.
Here, the label $\alpha = \mathrm{c,t}$ denotes the c- and t-OMS, respectively.
The optical mode is coupled with the mechanical vibration of the cavity mirror of the OMS or a specific OMS structure by
the optical pressure in the cavity and a shift of the resonant light frequency owing to the position modulation.
The mechanical vibration is described by $\hat{b}_\alpha , \hat{b}_\alpha^\dagger$.
Then, the Hamiltonian of the OMSs is written as follows~\cite{PhysRevA.51.2537,RevModPhys.86.1391}:
\begin{eqnarray}
\hat{H}_{\mathrm{OMS}} &=& \sum_{\alpha = {\rm c,t}} \left[
\left( \hbar \omega_\alpha \hat{a}_\alpha^{\dagger} \hat{a}_\alpha
+ \hbar \Omega_\alpha \hat{b}_\alpha^{\dagger} \hat{b}_\alpha \right) \right.\nonumber \\
&-& \left. \hbar g_\alpha \hat{a}_\alpha^{\dagger} \hat{a}_\alpha \left(\hat{b}_\alpha^{\dagger} + \hat{b}_\alpha \right) \right].
\label{eq:Homs}
\end{eqnarray}
Here, $\omega_\alpha$ and $\Omega_\alpha$ are the eigenfrequencies of the optical and mechanical modes, respectively.
$g_\alpha$ is the coupling strength of the optical mode and the mechanical vibration determined by the OMS structure.
We use the unit of $\hbar=1$ for simplicity in the following.
In this study, we consider identical OMSs, $\omega_{\rm c} = \omega_{\rm t} \equiv \omega_0$,
$\Omega_{\rm c} = \Omega_{\rm t} \equiv \Omega_0$, and $g_{\rm c} = g_{\rm t} \equiv g_0$, to highlight the significant mechanism.

The mechanical coupling between the OMSs via their substrate is given as
$\hat{H}_{\mathrm{coup}} = -K \left( \hat{b}_{\rm t}^\dagger \hat{b}_{\rm c}
+ \hat{b}^\dagger_{\rm c} \hat{b}_{\rm t} \right)$~\cite{PhysRevLett.111.073603,PhysRevX.5.031011},
where the coupling strength $K$ relates to the phonon transport in the substrate.
The strength $K$ is designed by the system fabrication.
In some recent fabrications~\cite{PhysRevLett.123.017402,PhysRevApplied.14.014041},
a mechanical coupling strength up to $K \lesssim \Omega_0$ has been achieved.
Note again that we assume that the OMSs do not communicate optically with each other.

In the proposed system, we apply a two-tone laser to the c-OMS with the resonant frequency of the optical mode, $\omega_0$,
and the blue- or red-detuned frequency by the phonon energy, $\omega_0 \pm \Omega_0$, while the t-OMS is irradiated by a monochromatic laser.
We describe the case of $\omega_0$ and $\omega_0 + \Omega_0$ irradiation on the c-OMS and $\omega_0 + \Omega_0$ irradiation on the t-OMS.
The Hamiltonian for the input lasers is as follows:
\begin{eqnarray}
\label{eq:Hlaser}
&\hat{H}_{\mathrm{drive}}&
= \left[ i \left( E_{{\rm c}, 0} e^{-i \omega_0 t } + E_{{\rm c}, +} e^{-i(\omega_0 + \Omega_0) t} \right) \hat{a}_{\rm c} \right. \nonumber\\
&&+ \left. i E_{{\rm t}, +} e^{-i(\omega_0 + \Omega_0) t} \hat{a}_{\rm t} \right] + \mathrm{h.c.},
\end{eqnarray}
where $E_{\alpha ,0}$ and $E_{\alpha ,+}$ are the strengths of the resonant and blue-detuned coherent driving fields, respectively.
These two plus one laser inputs are intended as a typical situation to generate phonons in the c-OMS and induce the Stokes scattering in the t-OMS by the generated phonons.
If we consider other situations for the laser incidence, the laser amplitude in Eq.\ (\ref{eq:Hlaser}) is replaced such as
$E_{{\rm t}, +} e^{-i(\omega_0 + \Omega_0) t} \to E_{{\rm t}, -} e^{-i(\omega_0 - \Omega_0) t}$.

The photons of the optical mode and the phonons of the mechanical vibration in the OMSs relax to the environment.
We introduce operators $\hat{c}_\alpha$ and $\hat{d}_\alpha$ to describe the relaxation of the photons and phonons, respectively.
The Hamiltonian of the environment with the interaction is
\begin{eqnarray}
& \hat{H}_{\mathrm{env}} &=\sum_{\alpha
= {\rm c,t}} \left[ \int d\omega \ \omega \hat{c}_\alpha^\dagger \hat{c}_\alpha
+ \int d\omega \ i \sqrt{\frac{\kappa}{2\pi}}
\left( \hat{a}_\alpha^\dagger \hat{c}_\alpha + \hat{c}_\alpha^\dagger \hat{a}_\alpha \right) \right]
\nonumber \\
&+& \hspace{-3mm} \sum_{\alpha = {\rm c,t}} \left[ \int d\Omega \ \Omega \hat{d}_\alpha^\dagger \hat{d}_\alpha
+ \int d\Omega \ i \sqrt{\frac{\gamma}{2\pi}}
\left( \hat{b}_\alpha^\dagger \hat{d}_\alpha + \hat{d}_\alpha^\dagger \hat{b}_\alpha \right) \right] \hspace{-1mm} ,
\label{eq:Henv}
\end{eqnarray}
where $\kappa$ and $\gamma$ are the damping constants of the optical and mechanical vibration modes, respectively.

The entire Hamiltonian is $\hat{H} = \hat{H}_{\mathrm{sys}} + \hat{H}_{\mathrm{env}}$ with
$\hat{H}_{\mathrm{sys}} = \hat{H}_{\mathrm{OMS}} + \hat{H}_{\mathrm{coup}} + \hat{H}_{\mathrm{drive}}$.
In this model, the photon in the t-OMS is {\it indirectly} coupled with the photon of the control OMS via the mechanical vibrations of the substrate.
The environment photons (phonons) described by $\hat{c}_\alpha^\dagger$ ($\hat{d}_\alpha^\dagger$) can enter the $\alpha$ OMS.
However, we assume that the $c$-photon ($d$-phonon) is absent in the initial state, and the relaxed photons (phonons) from the OMSs immediately move away.
Moreover, they do not go into the other OMSs.
This assumption corresponds to a low-noise environment at low temperature.
Then, we formulate the system $\hat{H}_{\mathrm{sys}}$ and treat the influence of the environment $\hat{H}_{\mathrm{env}}$
only as the relaxation in the Heisenberg equations (in Sec. III of Supplementary Materials).

We assume conventional cavity-quantum-electrodynamic (QED) systems to build up the present mechanically coupled OMSs.
The parameters for the energies of the cavity and mechanical vibration modes, coupling strengths,
and damping constants are typically
$\omega_0 \sim 200 \, \mathrm{THz}$ ($\lambda \sim 1550 \, \mathrm{nm}$),
$\Omega_0 \sim 1\, \mathrm{MHz} - 1\, \mathrm{GHz}$, $g_0/\Omega_0 \sim 10^{-4}$, $\kappa /\Omega_0 \sim 0.1$,
$\gamma /\Omega_0 \sim 10^{-4}$, and $K /\Omega_0 \sim 10^{-3} - 1$~\cite{PhysRevLett.123.017402}.
Hence, the cooperativity $C \equiv 4g_0^2 / (\gamma \kappa)$ and multiphoton cooperativity $C_{\mathrm{m}} \equiv \bar{n}_{\mathrm{cav}}C$,
which are important indices as a figure of merit of the QED systems~\cite{Clerk:2020aa,RevModPhys.86.1391}, are much smaller than unity.
$\bar{n}_{\mathrm{cav}}$ is the number of cavity photons.
It is worth noting that the proposed system provides detectable and noiseless Stokes signals, even for $C_{\mathrm{m}} \ll 1$ at the respective OMSs.

First, we observe the favorable condition for the mechanical components $\gamma$ and $K$ in a typical parameter regime of $\omega_0 \gg \Omega_0 > \kappa \gg g_0 ,\gamma ,K$.
We present the derivation and a detailed discussion in Sec.\ I of the Supplementary Materials; therefore, we only show the result here. 
Under a steady-state condition, the present system can be analytically solved, and we obtain the explicit expression of the expectation value
that explains the enhancement of the Stokes scattering component (Eq.\ (1) in Sec.\ I of the Supplementary Materials).
Examining this expression, we know that the optimal condition is given by $K \sim \gamma /2$ (see Fig.\ \ref{fig:model} in the Supplementary Materials).
This result attributes to that an increase in $K$ enhances the inelastic signal, while  a strong coupling of $K$ does not contribute to the enhancement of
$\left| \braket{\hat{a}_{{\rm t},0}} \right|$ owing to the energy level splitting of the phonons by $K$ (see Sec. I in the Supplementary Materials).

We discuss the efficiency of the enhancements by numerically calculating the output power spectrum at the t-OMS under the above-mentioned optimal condition.
For the numerical calculation, the system Hamiltonian $\hat{H}_{\mathrm{sys}}$ is transferred by a unitary operator,
$\hat{U}_2 = \exp\left[ i(\omega_0 + \Omega_0) \left( \sum_\alpha \hat{a}_\alpha^\dagger \hat{a}_\alpha \right) t \right]$
(for details, see Secs.\ II and III in the Supplementary Materials).

The output power spectrum of the optical mode in each OMS is evaluated using the Fourier transformation of the correlation function:
\begin{equation}
\mathcal{S}_\alpha (\omega) = \int \braket{\hat{a}_\alpha^\dagger (0) \hat{a}_\alpha (t)}e^{-i\omega t} dt.
\label{eq:power}
\end{equation}
$\braket{\hat{\mathcal{O}} (0)}$ denotes the time-averaged steady-state mean value of an arbitrary system operator $\hat{\mathcal{O}}(t)$.
This spectrum should be evaluated from the steady state (see Sec.\ V in the Supplementary Materials).

\begin{figure}[t]
\centering
\includegraphics[width=0.5\linewidth]{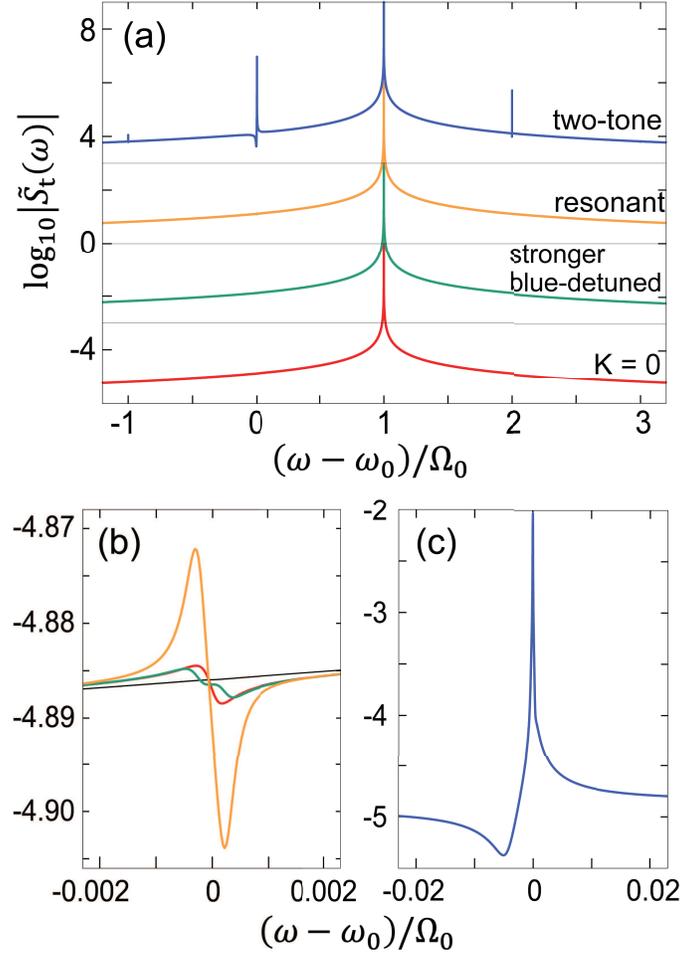}
\caption{
(a) Normalized output power spectra $|\tilde{\mathcal{S}}_{\rm t} (\omega)|$ of the target photons for
various conditions of the input laser beams at the c-OMS.
A monotone laser applied to the t-OMS is fixed at $\omega = \omega_0 + \Omega_0$.
The system parameters are as follows if the parameters are not designated:
$E_{\alpha,\xi}/\Omega_0 = 1.0$ ($\alpha = {\rm c,t}$ for the OMSs, $\xi = 0,+$), $\kappa/\Omega_0 = 0.23$,
$g_0/\Omega_0 = 2.4\times 10^{-4}$, $K/\Omega_0 = 2.35 \times 10^{-4}$, and $\gamma/\Omega_0 = 4.7\times 10^{-4}$.
The red, green, orange, and blue lines indicate the cases of no mechanical coupling ($K=0$), a monotone blue-detuned $\omega_0 + \Omega_0$ input,
a resonant $\omega_0$ input, and a two-tone ($\omega_0$ and $\omega_0 + \Omega_0$) input at the c-OMS, respectively.
The plotted curves are displayed using vertical shifts by 3 units to assist clear observation.
For the green line, the input laser power at the c-OMS is stronger ($E_{{\rm c},+}/\Omega_0=4.0$) to enhance phonon generation.
All spectra are normalized by their respective maximum value.
(b,c) Enlarged views of the Stokes scattering signals for the respective cases.
The line colors correspond to those in panel (a).
The vertical shift is not applied to (b) and (c).
The thin black line in (b) indicates a case of the resonant $\omega_0$ input and $K=0$,
which has a main peak at $\omega = \omega_0$; hence, the plot is shifted by $+1$ in the horizontal direction.
}
\label{fig:reslut_num}
\end{figure}

The evaluated output power spectrum at the t-OMS is shown in Fig.\ \ref{fig:reslut_num} for several input situations.
Here, we consider four cases: no mechanical coupling ($K=0$), a monotone blue-detuned $\omega_0 + \Omega_0$ strong input,
a resonant $\omega_0$ input, and a two-tone ($\omega_0$ and $\omega_0 + \Omega_0$) input at the c-OMS.
At the t-OMS, a blue-detuned monotone laser is applied in all cases.
In Fig.\ \ref{fig:reslut_num}(a), the spectrum exhibits sharp signals at $\omega \approx \omega_0 + (n+1) \Omega_0$ ($n$ is an integer).
Note that each spectrum is normalized by its maximum value, and the lines are plotted with vertical shifts.
The largest peaks for all the spectra at $\omega = \omega_0 + \Omega_0$ are attributed to the input laser frequencies.
When the OMSs are disconnected ($K=0$), the mechanical vibration at the t-OMS is induced only by the monotone input at the t-OMS.
Then, the inelastic-scattering signal induced by locally generated phonons is almost invisibly weak.
Figure \ref{fig:reslut_num}(b) displays an enlarged view that exhibits the weak signals at $\omega \approx \omega_0$.
The peak-and-dip height from a baseline output due to the $\omega_0 + \Omega_0$ input signal is of the order of $0.001$ in the figure.
Note that the plot in Fig.\ \ref{fig:reslut_num} is in the logarithmic scale.
Hence, a measurement of this Stokes signal requires an extremely high-quality OMS.
At $\omega \approx \omega_0 + 2 \Omega_0$, an anti-Stokes component might be present.
However, the signal is not identified in the spectrum.
For an incident resonant laser ($\omega = \omega_0$) at $K=0$ (see the black line in Fig.\ 2(b)),
the main peak produced by the incident laser is especially large, and the Stokes and anti-Stokes signals at
$\omega \approx \omega_0 \mp \Omega_0$, are completely masked by the baseline, owing to the input (no peak in the black line). 
Hence, the blue- or red-detuned laser incidence is appropriate to observe the inelastic signals.

When $K \ne 0$, the phonons generated at the c-OMS contribute to inelastic scattering at the t-OMS.
However, if the input at the c-OMS is monotone, the inelastic scattering signal at the t-OMS is negligible
even if the input is four times stronger than the blue-detuned input (green) or the resonant frequency input (orange lines) at the c-OMS.
Thus, phonon generation is ineffective under a monotone laser incidence.
In Fig.\ \ref{fig:reslut_num}(b), the Stokes scattering signal for the resonant monotone input at the c-OMS (orange) is
larger than that for the blue-detuned monotone input (green).
This indicates that phonon generation is efficient when the input laser is resonant
although local Stokes and anti-Stokes signals by the locally generated phonon ($K=0$ case) are masked when the input is resonant.
Therefore, the present nonlocal scheme is incredibly useful for detecting phonon generation and inelastic scattering because it avoids noise caused by laser incidence.

We discuss the drastic enhancement of the phonon generation and inelastic scatterings beyond those caused by the monotone input mentioned above.
The key ingredient is a two-tone input laser applied only to the c-OMS.
We examine the $\omega_0$ and $\omega_0 + \Omega_0$ two-tone input (blue line), as shown in Fig.\ \ref{fig:reslut_num}(a) and (c).
The two-tone input ``activates'' the c-OMS and nonlocally results in enormously enlarged inelastic scattering signals at the t-OMS.
The Stokes scattering at $\omega = \omega_0$ ($n=-1$) is significantly enlarged, owing to the optical cavity mode.
The enhancement from the baseline due to the input is approximately $10^3$ (see Fig.\ \ref{fig:reslut_num}(c)).
Here, we consider a signal-to-noise (SN) ratio.
The intensity of the noise caused by the input laser is approximated to be of the same order as that of the output spectrum, owing to the baseline of the input laser.
Then, the ratio between the peak height and the baseline gives a qualitative estimation of the SN ratio.
Let us call this the $\eta$-factor, $\eta \equiv ({\rm peak \ height})/({\rm based \ line})$.
The $\eta$-factor for the local Stokes scattering induced by the blue-detuned laser with $K=0$ (red) is
$\eta \approx 10^{-4.8845}/10^{-4.8860} = 10^{0.0015}$ from Fig.\ \ref{fig:reslut_num}(b).
For the nonlocal Stokes scattering by the resonant monotone laser (orange), $\eta \approx 10^{0.014}$.
This suggests the important insight that a nonlocal signal would result in a large SN ratio with noise suppression.
However, the enhancement is still weak.
In our proposed scheme under the two-tone laser, the nonlocal Stokes signal indicates $\eta \approx 10^{2.75}$ (blue line),
which is considerably larger than that in other situations.
In addition to enhancement (peak), the spectrum indicates suppression (dip).
Such a peak-dip spectrum is attributed to the Fano resonance between the mechanical vibration mode and the baseline spectrum of the photons in the cavity.

\begin{figure}[t]
\centering
\includegraphics[width=0.5\linewidth]{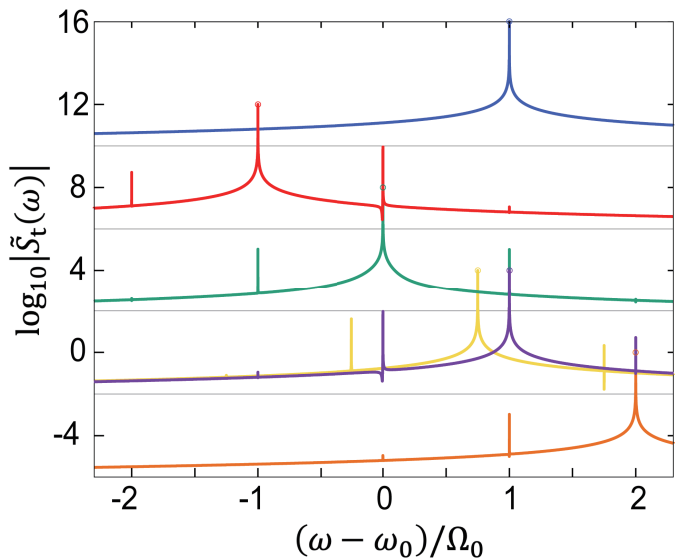}
\caption{
Normalized output power spectra $|\tilde{\mathcal{S}}_{\rm t} (\omega)|$ of the target photons
when two-tone and monotone lasers are applied to the c- and t-OMSs, respectively.
For the efficient generation of phonons, $\omega_0$ and $\omega_0 + \Omega_0$ two-tone input is considered with
a monotone input of $\omega = \omega_0 + 2\Omega_0$ (orange), $\omega_0 + \Omega_0$ (purple),
$\omega_0$ (green), and $\omega_0 - \Omega_0$ (red lines) at the t-OMS.
A case of $\omega_0 - \Omega_0$ and $\omega_0 + \Omega_0$ input at the c-OMS and $\omega_0 + \Omega_0$ input at the t-OMS is plotted (blue line).
The respective curves are plotted with lateral offsets of 3 units to assist with clear observation.
The other parameters are the same as those in Fig.\ \ref{fig:reslut_num}.
Hence, the purple line is identical to the blue line in Fig.\ \ref{fig:reslut_num}.
A case of $\omega_0 - (1/2) \Omega_0$ and $\omega_0 + (1/2) \Omega_0$ input at the c-OMS and
$\omega_0 + \Omega_0$ input at the t-OMS is also shown with a horizontal offset by $-0.25$ (yellow line).
All spectra are normalized by the respective maximum value.
Each maximum point, which value is unity, is marked by a blank circle.
}
\label{fig:TC}
\end{figure}

Next, we examine the dependence on the frequencies of the two-tone and monotone input lasers.
In Fig.\ \ref{fig:TC}, $|\tilde{\mathcal{S}}_{\rm t} (\omega)|$ indicates the enhanced signals of the nonlocal inelastic scatterings.
The input frequency to the t-OMS is changed to $\omega_0 + 2\Omega_0$ (doubly blue-detuned, orange),
$\omega_0 + \Omega_0$ (blue-detuned, purple), $\omega_0$ (resonant, green), and $\omega_0 -\Omega_0$ (red-detuned, red lines) with
fixed $\omega_0$ and $\omega_0 + \Omega_0$ two-tone input at the c-OMS.
The Stokes scattering signal reaches a maximum when the blue-detuned input is at the t-OMS,
whereas the red-detuned input maximizes the anti-Stokes scattering.
The resonant input at the t-OMS (green line) is not an optimal condition.
The maximized condition of the Stokes (anti-Stokes) scattering is found because the Stokes (anti-Stokes) shift is on the optical cavity mode of the t-OMS.
Then, the peak height of the Stokes (anti-Stokes) signal is increased to $10^{3}$ from
the baseline of the incident frequency (see purple (red) line at $\omega = \omega_0$).
For the doubly blue-detuned input (orange line), the first-order inelastic scatterings are enhanced,
whereas the second-order inelastic scattering caused by two phonons does not exhibit a large signal in the present consideration,
even for the $\omega = \omega_0$ optical cavity mode.
These enhancements of the inelastic scattering signals are supported by the efficient phonon generation at the c-OMS.
Efficient generation is also caused by the $\omega_0 - \Omega_0$ and $\omega_0$ two-tone input.
Then, the spectra are overlapped with the shown results.

Two-phonon generation at the c-OMS is ineffective when the $\omega_0 - \Omega_0$ and $\omega_0 + \Omega_0$ two-tone laser is applied (blue line).
In this case, the inelastic scattering signal is not found in the spectrum.
We examine the $\omega_0 - \Omega_0/2$ and $\omega_0 + \Omega_0/2$ two-tone laser incidence (yellow line with a small horizontal shift).
Then, one-phonon generation is effective, and the inelastic scattering signal is sufficiently large.
Its peak height is more than $10^{2}$ from the baseline (see the peak of the yellow line at $\omega \approx \omega_0$).
Hence, the two-tone input with an $\Omega_0$ frequency difference is one of the required conditions to ``activate'' the c-OMS,
whereas an optical resonant condition for the cavity mode is not strictly required.
Additionally, coherent phonon transport between the OMSs is required, and
the difference between the mechanical vibration modes of the OMSs should be small, $|\Omega_{\rm c} - \Omega_{\rm t}| < \sqrt{\gamma^2 +K^2}$.

In conclusion, nonlocal inelastic scattering by remote and mechanically coupled OMSs generates noiseless and pure inelastic signals.
The two-tone laser beams ``activate'' the c-OMS to generate phonons that affords the nonlocal inelastic signals from the t-OMS. 
The frequency difference between the two tones corresponding to the mechanical vibration mode significantly enhances inelastic signals by several orders of magnitude.
If the difference departed from the mechanical vibration mode, the inelastic signals disappeared sensitively. 
In addition, if a molecule or gas is adsorbed into one OMS and, the mechanical vibration mode is changed,
the inelastic scattering signal is weakened (enlarged), owing to the suppression (enhancement) of phonon transport,
which leads to a sensitive detection of the molecule or gas.
Thus, our scheme can be applied to remote (noiseless), robust, and accurate sensors.

The present conclusions are derived from our specific model of remote-coupled OMSs.
However, the physics discussed in this paper might have a general aspect toward nonlocal inelastic scattering.
The key concept is nonlocal control of inelastic scattering by two-plus-one color lasers,
which has considerable potential to open optical phononics, for example, in local phonon generation by a focused laser under a wide-area reference light.

This work is supported in part by JSPS KAKENHI 18K13484 and 18H01151.

\appendix

\section{Analytical solution}
\label{sec:analytical_solution}
\begin{figure}[t]
\centering
\includegraphics[width=\linewidth]{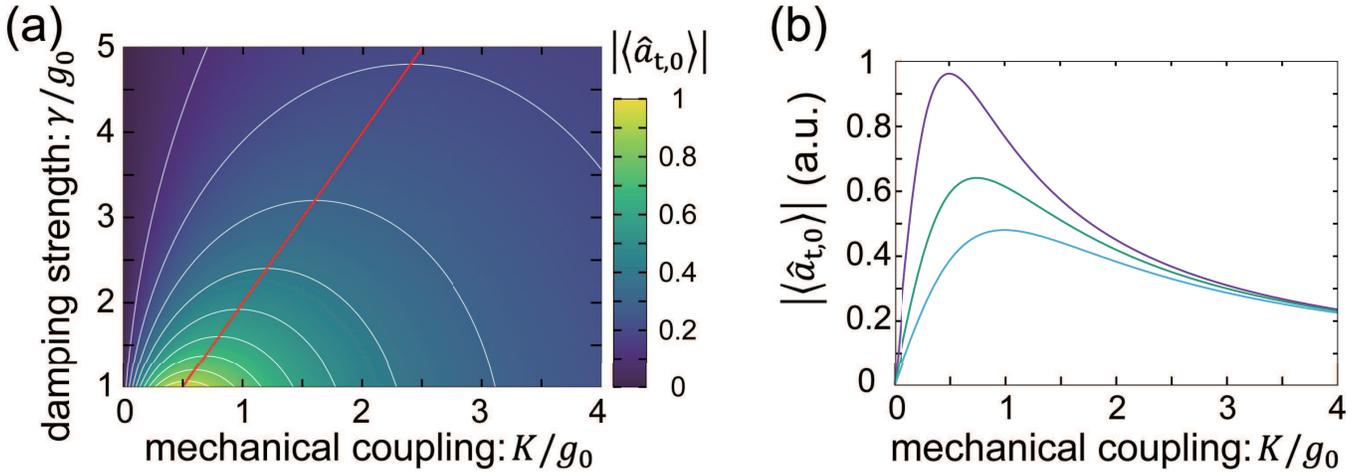}
\caption{
(a) Expectation value $|\braket{\hat{a}_{{\rm t},0}}|$, which exhibits the enhancement of Stokes scattering from the t-OMS
in the plane of the mechanical coupling strength $K$ and mechanical damping constant $\gamma$.
The parameters of the OMSs are $g_0 = 2.4 \times 10^{-4} \Omega_0$ and $\kappa=0.23 \Omega_0$.
The incident laser intensities parametrized by $E_{\alpha, \xi}$ in the Hamiltonian are set to $|E_{\rm c,0}| = |E_{\rm c,+}| = |E_{\rm t,+}| = \Omega_0$.
Note that the phase factor of $E_{\alpha, \xi}$ is irrelevant to $|\braket{\hat{a}_{{\rm t},0}}|$.
Thin white lines denote contour lines with an interval of $0.1$.
The thick red line denotes the maximum position in $K$ when $\gamma$ is fixed, which indicates the optimal condition for inelastic scattering enhancement.
(b) Cross-sectional plot of $|\braket{\hat{a}_{{\rm t},0}}|$ in (a) at $\gamma /g_0 = 1.0$ (purple), $1.5$ (green), and $2.0$ (light blue lines).
}
\label{fig:result_analytical}
\end{figure}

Here, we examine an analytical solution to discuss the enhancement of Stokes scattering.
We focus on the mechanical components $\gamma$ and $K$ in a typical parameter regime of $\omega_0 \gg \Omega_0 > \kappa \gg g_0 ,\gamma ,K$.
For the system Hamiltonian $\hat{H}_{\mathrm{sys}}$ described in the main text with mechanically coupled optomechanical systems (OMSs),
we apply a unitary transformation with respect to the operator,
$\hat{U}_1 (t) = \mathrm{exp} \left[ i \sum_{\alpha = {\rm c,t}}
\left(\omega_0 \hat{a}_\alpha^\dagger \hat{a}_\alpha + \Omega_0 \hat{b}_\alpha^\dagger \hat{b}_\alpha \right) t\right]$,
to formulate the Heisenberg equations of motion in a simple description (see Sec.\ \ref{appen:Utrans}).
Here, the indices c and t represent the controller OMS (c-OMS) and target OMS (t-OMS), respectively.
Assuming the condition $\kappa \ll \Omega_0$, where inelastically generated light spectra are separated from the resonant signal,
we take an ansatz for the photons~\cite{PhysRevA.89.063805,PhysRevA.78.062303}:
$\hat{a}_{\rm c} = \hat{a}_{{\rm c},0} + \hat{a}_{{\rm c},+} e^{-i\Omega_0 t} + \hat{a}_{{\rm c},-} e^{i\Omega_0 t}$, and
$\hat{a}_{\rm t} = \hat{a}_{{\rm t},0} + \hat{a}_{{\rm t},+} e^{-i\Omega_0 t}$.
The labels $0$, $+$, and $-$ represent the resonant, blue-detuned, and red-detuned components, respectively.
Here, the higher-order inelastic scattered photons are not considered.
We assume that for the two-tone and monotone inputs to the c- and t-OMSs, respectively,
the operators $\hat{a}_{{\rm c},0}$, $\hat{a}_{{\rm c},+}$, and $\hat{a}_{{\rm t},+}$ in the Heisenberg equations are constant.
Then, the equations of motion for $\hat{a}_{{\rm c},-}$, $\hat{a}_{{\rm t},0}$, $\hat{b}_{\rm c}$, and $\hat{b}_{\rm t}$ are linearized.
This set of inhomogeneous linear equations of operators is analytically solvable. (See, Eqs.(\ref{eq:am-_2})--(\ref{eq:bs}).)
When all photons and phonons are in the steady-states, $d\braket{\hat{a}_{\alpha,\xi}} /dt=0$ and $d\braket{\hat{b}_\alpha } /dt=0$, we obtain
\begin{equation}
\label{eq:steadyas0}
\braket{\hat{a}_{{\rm t},0}} =
\frac{i\bar{g}_{\rm t} \bar{g}_{\rm cc}^* K}{\zeta |\bar{g}_{\rm t}|^{2} - \frac{\kappa}{2} \left( K^2 + \zeta \gamma/2 \right) },
\end{equation}
with $\zeta = \frac{\gamma}{2} - \frac{|g_{\rm c}|^2}{\kappa/2 + i\Omega_0}$,
$\bar{g}_{\rm cc} = g_0 \bar{a}_{{\rm c},0}^* \bar{a}_{{\rm c},+}$,
$\bar{g}_{\rm c } = g_0 \bar{a}_{{\rm c},0}$, and $\bar{g}_{\rm t } = g_0 \bar{a}_{{\rm t},+}$.
(see the detailed derivation in Sec.\ \ref{appen:Utrans}).

Figure \ref{fig:result_analytical}(a) illustrates the absolute value of $\braket{\hat{a}_{{\rm t},0}}$ in the $\gamma$-$K$ plane.
For a fixed $K$, $\left| \braket{\hat{a}_{{\rm t},0}} \right|$ is monotonically reduced by the damping constant $\gamma$ of the mechanical vibration.
When $\gamma$ is fixed, on the other hand, there is an optimal condition, shown in Fig.\ \ref{fig:result_analytical}(b).
The condition can be approximated when the numerator of $d \braket{\hat{a}_{{\rm t},0}}/dK$ is equal to zero, $K^2 \sim \frac{\gamma^2}{4}$.
At $K=0$, an inelastic scattering signal is not observed because phonon transport is absent.
An increase in $K$ enhances the inelastic signal.
However, the stronger coupling of $K$ does not contribute to the enhancement of $\left| \braket{\hat{a}_{{\rm t},0}} \right|$.
This is attributed to the energy shift of the mechanical vibration modes by $K$. 
Hence, the two-tone input laser with $\omega_0$ and $\omega_0 + \Omega_0$ frequencies,
which correspond to the resonant and blue-detuned ``eigen'' frequencies of the isolated OMS, cannot generate phonons effectively.
The damping constant $\gamma$ indicates the energy-width of the phonon spectrum.
Thus, the optimal condition is given as $K \sim \gamma /2$.
If the ``eigen'' frequencies of the input laser are appropriately tuned for a large $K$, the enhancement factor can be larger.
However, this condition becomes system-dependent, which is undesirable for applications.

\section{Numerical calculation: Fourier spectrum}
\label{sec:numerical_solution}
To discuss the enhancement efficiency, we examine a numerical calculation of the output power spectrum at the t-OMS.
For the numerical calculation, the system Hamiltonian is transferred by another unitary operator:
$\hat{U}_2 = \exp\left[ i(\omega_0 + \Omega_0) \left( \sum_\alpha \hat{a}_\alpha^\dagger \hat{a}_\alpha \right) t \right]$.
(see Sec.\ \ref{appen:Utrans} for details).

The output power spectrum of the optical mode in each OMS is evaluated using the Fourier transformation of the correlation function:
\begin{equation}
\mathcal{S}_\alpha (\omega) = \int \braket{\hat{a}_\alpha^\dagger (0) \hat{a}_\alpha (t)}e^{-i\omega t} dt.
\label{eq:power}
\end{equation}
Here, $\braket{\hat{\mathcal{O}} (0)}$ denotes the time-averaged steady-state mean value of an arbitrary system operator $\hat{\mathcal{O}}(t)$.
The time evolution follows the equations of motion (\ref{eq:Heisenberg21})--(\ref{eq:Heisenberg23}), which consist of multiple operators.
After taking the expectation of the operators, we adopt the cluster expansion method to
solve the equations (see Sec.\ref{appen:cluster} and Ref. [\onlinecite{PhysRevA.78.022102}].)
The correlation function is divided into two terms, e.g.,
\begin{equation}
\braket{\hat{a}_\alpha^\dagger (0) \hat{a}_\alpha (t)}
= \braket{\hat{a}_\alpha^\dagger (0)} \braket{\hat{a}_\alpha (t)}
+ \Delta \braket{\hat{a}_\alpha^\dagger (0) \hat{a}_\alpha (t)}.
\label{eq:correlation}
\end{equation}
The second term on the right-hand-side represents the quantum correlation.
The time evolution of $\braket{\hat{a}_\alpha^\dagger (0) \hat{a}_\alpha (t)}$ also follows Eqs.\ (\ref{eq:Heisenberg21})--(\ref{eq:Heisenberg23}).
The third-order terms, such as $\Delta \braket{\hat{a}_\alpha (0) \hat{a}_\alpha^\dagger (t) \hat{a}_\alpha (t)}$, and
the higher-order terms are deduced in the equations of motion.
The order of the operators increases in the inductive self-consistent treatment of the equations.
There is a cut-off in the weak coupling regime for the respective OMSs,
where $C = 4g_0^2 / (\gamma \kappa) \ll 1$ is satisfied, and we can neglect the third- and higher-order correlations.

\section{Unitary transformation}
\label{appen:Utrans}
In this section, we describe two unitary transformations of the system Hamiltonian $\hat{H}_{\mathrm{sys}}$ in the main text.
These are utilized in Secs.\ \ref{sec:analytical_solution} and \ref{sec:numerical_solution}.

First, we consider a unitary transformation with respect to a rotating frame, as follows:
\begin{equation}
\hat{H}^\prime_{\mathrm{sys}} (t) = \hat{U}_1 (t) \hat{H}_{\mathrm{sys}} \hat{U}_1^\dagger (t)
-i \hat{U}_1 (t) \frac{\partial}{\partial t} \hat{U}_1^\dagger (t),
\end{equation}
with the unitary operator
\begin{equation}
\hat{U}_1 (t) = \mathrm{exp} \left[ i \sum_\alpha \left(\omega_0 \hat{a}_\alpha^\dagger \hat{a}_\alpha
+ \Omega_0 \hat{b}_\alpha^\dagger \hat{b}_\alpha \right) t\right].
\label{eq:U1}
\end{equation}
Then, we obtain
\begin{eqnarray}
\label{eq:Hanaly}
\hat{H}^\prime_{\mathrm{sys}} &=& -g_0 \sum_{\alpha = {\rm c,t}} \hat{a}_\alpha^\dagger \hat{a}_\alpha
( e^{i\Omega_0 t} \hat{b}_\alpha^\dagger + e^{-i\Omega_0 t} \hat{b}_\alpha )
- K \left( \hat{b}_{\rm t}^\dagger \hat{b}_{\rm c} + \hat{b}_{\rm c}^\dagger \hat{b}_{\rm t} \right),
\nonumber \\
& & \hspace{5mm} + \left[ i \left( E_{{\rm c},0} + E_{{\rm c},+} e^{-i\Omega_0 t} \right) \hat{a}_{\rm c}
+ iE_{{\rm t},+}e^{-i\Omega_0 t} \hat{a}_{\rm t} + \mathrm{h.c.} \right].
\end{eqnarray}
Here, we omit the description ``$(t)$'' of the operators for simplicity.
Note that this unitary transformation is available for different input laser frequencies.
For the transferred Hamiltonian (\ref{eq:Hanaly}), the Heisenberg equations of motion for the operators are
\begin{eqnarray}
\frac{d}{dt} \hat{a}_{\rm c} &=& -\frac{\kappa}{2} \hat{a}_{\rm c}
+ i g_0 \hat{a}_{\rm c} (e^{i\Omega_0 t} \hat{b}_{\rm c}^\dagger + e^{-i\Omega_0 t} \hat{b}_{\rm c})
+ E_{{\rm c},0} + e^{-i\Omega_0 t} E_{{\rm c},+},
\label{eq:am_analytical} \\
\frac{d}{dt} \hat{a}_{\rm t} &=& -\frac{\kappa}{2} \hat{a}_{\rm t}
+ i g_0 \hat{a}_{\rm t} (e^{i\Omega_0 t} \hat{b}_{\rm t}^\dagger + e^{-i\Omega_0 t} \hat{b}_{\rm t})
+ e^{-i\Omega_0 t} E_{{\rm t},+},
\label{eq:as_analytical} \\
\frac{d}{dt}\hat{b}_\alpha &=& -\frac{\gamma}{2}\hat{b}_\alpha
+ i g_{0} \hat{a}_\alpha \hat{a}_\alpha^\dagger e^{i\Omega_{0} t} + i K\hat{b}_{\bar{\alpha}}.
\label{eq:bm_analytical}
\end{eqnarray}
The first terms on the right-hand side of these equations arise from the coupling between the system and the environment~\cite{PhysRevA.31.3761}.
$\bar{\alpha}$ in Eq.\ (\ref{eq:bm_analytical}) represents $\bar{\rm c} = {\rm t}$ and $\bar{\rm t} = {\rm c}$.

Let us formulate the analytical solution in Sec.\ \ref{sec:analytical_solution}.
In this study, we assume that the relaxation of the optical cavity mode in the OMSs is much smaller than the mechanical vibration mode, $\kappa \ll \Omega_0$.
Then, the side-band spectra caused by the inelastic scattering of light are well separated from the main spectrum owing to the input laser.
For this situation, we take the following ansatz~\cite{PhysRevA.89.063805,PhysRevA.78.062303}:
\begin{eqnarray}
\label{eq:ansatz_am}
\hat{a}_{\rm c} &=& \hat{a}_{{\rm c},0} + \hat{a}_{{\rm c},+} e^{-i\Omega_0 t} + \hat{a}_{{\rm c},-} e^{i\Omega_0 t},
\\
\label{eq:ansatz_as}
\hat{a}_{\rm t} &=& \hat{a}_{{\rm t},0} + \hat{a}_{{\rm t},+} e^{-i\Omega_0 t}
\end{eqnarray}
by introducing the photon operators $\hat{a}_{\alpha,0}$, $\hat{a}_{\alpha,+}$, and $\hat{a}_{\alpha,-}$.
The labels $0$, $+$, and $-$ represent the resonant, blue-detuned, and red-detuned components, respectively.
Here, the higher-order inelastic scattered photons $\hat{a}_{\alpha ,\pm n} e^{\mp n\Omega_0 t}$ ($n \ge 2$) are neglected
because their energy is far from the energy of the OMS cavity mode.
For the t-OMS, we do not consider the $\hat{a}_{{\rm t},-}e^{i\Omega_0 t}$ term
because this non-resonant Fourier component is far from the input frequency $\omega_0 + \Omega_0$,
while the $\hat{a}_{{\rm c},-}e^{i\Omega_0 t}$ term is considered because it is a ``neighboring'' component of the input frequency $\omega_0$ in the c-OMS.

Substituting Eq.\ (\ref{eq:ansatz_am}) into the Heisenberg equation (\ref{eq:am_analytical}), we obtain the following:
\begin{eqnarray}
\label{eq:am0}
\frac{d}{dt} \hat{a}_{{\rm c},0} &=& -\frac{\kappa}{2} \hat{a}_{{\rm c},0}
+ ig_0 (\hat{a}_{{\rm c},+} \hat{b}_{\rm c}^\dagger + \hat{a}_{{\rm c},-} \hat{b}_{\rm c}) + E_{{\rm c},0},
\\
\label{eq:am+}
\frac{d}{dt} \hat{a}_{{\rm c},+} &=& \left(i\Omega_0 - \frac{\kappa}{2} \right) \hat{a}_{{\rm c},+}
+ ig_0 \hat{a}_{{\rm c},0} \hat{b}_{\rm c} + E_{{\rm c},+},
\\
\label{eq:am-}
\frac{d}{dt} \hat{a}_{{\rm c},-} &=& \left(-i\Omega_0 - \frac{\kappa}{2} \right) \hat{a}_{{\rm c},-}
+ ig_0 \hat{a}_{{\rm c},0} \hat{b}_{\rm c}^\dagger,
\end{eqnarray}
and the substitution of Eqs.\ (\ref{eq:ansatz_as}) into (\ref{eq:as_analytical}) gives
\begin{eqnarray}
\label{eq:as0}
\frac{d}{dt} \hat{a}_{{\rm t},0} &=& -\frac{\kappa}{2} \hat{a}_{{\rm t},0}
+ ig_0 \hat{a}_{{\rm t},+} \hat{b}_{\rm t}^\dagger,
\\
\label{eq:as+}
\frac{d}{dt} \hat{a}_{{\rm t},+} &=& \left(i\Omega_0 - \frac{\kappa}{2} \right) \hat{a}_{{\rm t},+}
+ ig_0 \hat{a}_{{\rm t},0} \hat{b}_{\rm t} + E_{{\rm t},+}.
\end{eqnarray}
Let us consider the expectation values of operators in the steady state when sufficient strong lasers are applied,
where the states are described by the coherent states.
For the two-tone and monotone laser incidences at the c- and t-OMSs, respectively,
the operators in the equations of motion (\ref{eq:am0}), (\ref{eq:am+}), and (\ref{eq:as+}) are replaced with
$\hat{a}_{\alpha, \xi} \to \braket{\hat{a}_{\alpha, \xi}} = \bar{a}_{\alpha, \xi}$ ($\alpha = {\rm c,t}$ and $\xi = 0,\pm$).
Then, we obtain
\begin{eqnarray}
\bar{a}_{{\rm c},0} &=& \frac{E_{{\rm c},0}}{\kappa /2},
\\
\bar{a}_{{\rm c},+} &=& \frac{E_{{\rm c},+}}{\kappa /2 - i\Omega_0},
\\
\bar{a}_{{\rm t},+} &=& \frac{E_{{\rm t},+}}{\kappa /2 - i\Omega_0},
\end{eqnarray}
when $g_0 \ll E_{\alpha,\xi},\kappa ,\Omega_0$, which is a typical condition~\cite{PhysRevA.89.063805}.
These input photons become steady states fast if the incident light intensities are sufficiently strong.
On the other hand, the induced side-band photons and phonons require a longer time-scale to achieve a steady state compared with the input photons.
The side-band photons and phonons adhere to the following equations of motion:
\begin{eqnarray}
\label{eq:am-_2}
\frac{d}{dt} \hat{a}_{{\rm c},-} &=& \left( -i\Omega_0 - \frac{\kappa}{2} \right) \hat{a}_{{\rm c},-}
+ i \bar{g}_{\rm c} \hat{b}_{\rm c}^\dagger,
\\
\label{eq:as0_2}
\frac{d}{dt} \hat{a}_{{\rm t},0} &=& -\frac{\kappa}{2} \hat{a}_{{\rm t},0}
+ i \bar{g}_{\rm t} \hat{b}_{\rm t}^\dagger,
\\
\label{eq:bm}
\frac{d}{dt} \hat{b}_{\rm c} &=& -\frac{\gamma}{2} \hat{b}_{\rm c} + iK \hat{b}_{\rm t}
+ i\bar{g}_{\rm c} \hat{a}_{{\rm c},-}^\dagger + i \bar{g}_{\rm cc},
\\
\label{eq:bs}
\frac{d}{dt} \hat{b}_{\rm t} &=& -\frac{\gamma}{2} \hat{b}_{\rm t} + iK \hat{b}_{\rm c}
+ i \bar{g}_{\rm t} \hat{a}_{{\rm t},0}^\dagger,
\end{eqnarray}
where we define the effective coupling constants,
\begin{eqnarray}
\bar{g}_{\rm cc} &=& g_0 \bar{a}_{{\rm c},0}^* \bar{a}_{{\rm c},+}, \\
\bar{g}_{\rm c } &=& g_0 \bar{a}_{{\rm c},0}, \\
\bar{g}_{\rm t } &=& g_0 \bar{a}_{{\rm t},+}.
\end{eqnarray}
Eqs.\ (\ref{eq:am-_2})--(\ref{eq:bs}) provide a set of inhomogeneous linear equations of operators.
The solutions are given analytically for the case in which the input laser intensities are constant.
When all photons and phonons become steady states, $d\braket{\hat{a}_{\alpha,\xi}} /dt=0$ and $d\braket{\hat{b}_\alpha } /dt=0$,
we obtain Eq.\ (\ref{eq:steadyas0}) with
\begin{equation}
\zeta = \frac{\gamma}{2} - \frac{|g_{\rm c}|^2}{\kappa/2 + i\Omega_0}.
\end{equation}

Next, for the numerical calculations described in Sec.\ \ref{sec:numerical_solution}, we consider another unitary transformation,
\begin{equation}
\hat{U}_2 = \exp\left[ i(\omega_0 + \Omega_0) \left( \sum_\alpha \hat{a}_\alpha^\dagger \hat{a}_\alpha \right) t \right],
\label{eq:U2}
\end{equation}
for the system Hamiltonian $\hat{H}_\mathrm{sys}$:
\begin{eqnarray}
\label{eq:hamiltonian_numerical}
\hat{H}_\mathrm{sys}^{''} &=& \hat{U}_2 \hat{H}_\mathrm{sys} \hat{U}_2^\dagger
-i \hat{U}_2 \frac{\partial}{\partial t} \hat{U}_2^\dagger
\nonumber \\
&=& \sum_{\alpha = {\rm c,t}} \left[ -\Omega_0 \hat{a}_\alpha^\dagger \hat{a}_\alpha + \Omega_0 \hat{b}_\alpha^\dagger \hat{b}_\alpha
 - g_0 \hat{a}_\alpha^\dagger \hat{a}_\alpha \left(\hat{b}_\alpha^\dagger + \hat{b}_\alpha \right) \right]
 - K \left( \hat{b}_{\rm c} \hat{b}_{\rm t}^\dagger + \hat{b}^\dagger_{\rm c} \hat{b}_{\rm t} \right) \nonumber\\
& & + \left[i \left( E_{{\rm c},0} e^{i \Omega_0 t} + E_{{\rm c},+} \right) \hat{a}_{\rm c}
 + iE_{{\rm t},+}\hat{a}_{\rm t} + \mathrm{h.c.} \right].
\end{eqnarray}
The Heisenberg equations for the operators are as follows:
\begin{eqnarray}
\frac{d}{dt} \hat{a}_{\rm c} &=& \left(i\Omega_0 - \frac{\kappa}{2}\right) \hat{a}_{\rm c}
+ i g_0 \hat{a}_{\rm c} \left( \hat{b}_{\rm c}^\dagger + \hat{b}_{\rm c} \right)
+ E_{{\rm c},0} e^{i\Omega_0 t} + E_{{\rm c},+},
\label{eq:Heisenberg21} \\
\frac{d}{dt} \hat{a}_{\rm t} &=& \left(i\Omega_0 - \frac{\kappa}{2}\right) \hat{a}_{\rm t}
+ i g_0 \hat{a}_{\rm t} \left(\hat{b}_{\rm t}^\dagger + \hat{b}_{\rm t} \right)
+ E_{{\rm t},+},
\label{eq:Heisenberg22} \\
\frac{d}{dt}\hat{b}_\alpha &=& \left(-i \Omega_0 - \frac{\gamma}{2}\right) \hat{b}_\alpha
+ i g_0 \hat{a}_\alpha^\dagger \hat{a}_\alpha + i K\hat{b}_{\bar{\alpha}}.
\label{eq:Heisenberg23}
\end{eqnarray}
Here, $\bar{\alpha}$ in (\ref{eq:Heisenberg23}) indicates that $\bar{\rm c} = {\rm t}$ and $\bar{\rm t} = {\rm c}$.
Note that this transformation applies to a rotating frame with a $\omega_0 + \Omega_0$ laser.
Then, the factors in the Heisenberg equations become time-independent, except for the third term on the right-hand side in
Eq.\ (\ref{eq:Heisenberg21}), which describes the $\omega_0$ input laser.
These transformations and Heisenberg equations provide an advantage for the numerical discussion.

\section{Cluster expansion}
\label{appen:cluster}
In this section, we explain a cluster expansion method~\cite{PhysRevA.78.022102},
which is applied to our numerical calculation in Eqs.\ (\ref{eq:Heisenberg21})--(\ref{eq:Heisenberg23}).
This method is a conventional treatment in quantum optics.
The cluster expansions for arbitrary operators, $\hat{X},\hat{Y},\hat{Z}$ are given as follows:
\begin{eqnarray}
\label{eq:doublet}
\braket{\hat{X} \hat{Y}}
&=& \braket{\hat{X}} \braket{\hat{Y}} + \Delta\braket{\hat{X} \hat{Y}},
\\
\label{eq:triplet}
\braket{\hat{X} \hat{Y} \hat{Z}}
&=& \braket{\hat{X}} \braket{\hat{Y}} \braket{\hat{Z}}
+ \braket{\hat{X}} \Delta \braket{\hat{Y} \hat{Z}} + \braket{\hat{Y}} \Delta \braket{\hat{X} \hat{Z}}
+ \braket{\hat{Z}} \Delta \braket{\hat{X} \hat{Y}}
\nonumber \\
& & + \Delta \braket{\hat{X}\hat{Y}\hat{Z}}.
\end{eqnarray}
Terms such as $\Delta \braket{\hat{X}\hat{Y}}$ are called doublets, which originate purely from quantum correlations.
Here, we neglect the triplet correlation terms $\Delta \braket{\hat{X} \hat{Y} \hat{Z}}$ in a weak coupling regime.
In this study, the weak coupling regime corresponds to a small cooperativity $C = 4g_0^2/(\gamma \kappa) \ll 1$.
Note that for a single operator, $\Delta \braket{\hat{a}_\alpha} = \Delta \braket{\hat{b}_\alpha} = 0$.

\begin{figure}[t]
\centering
\includegraphics[width=0.9\linewidth]{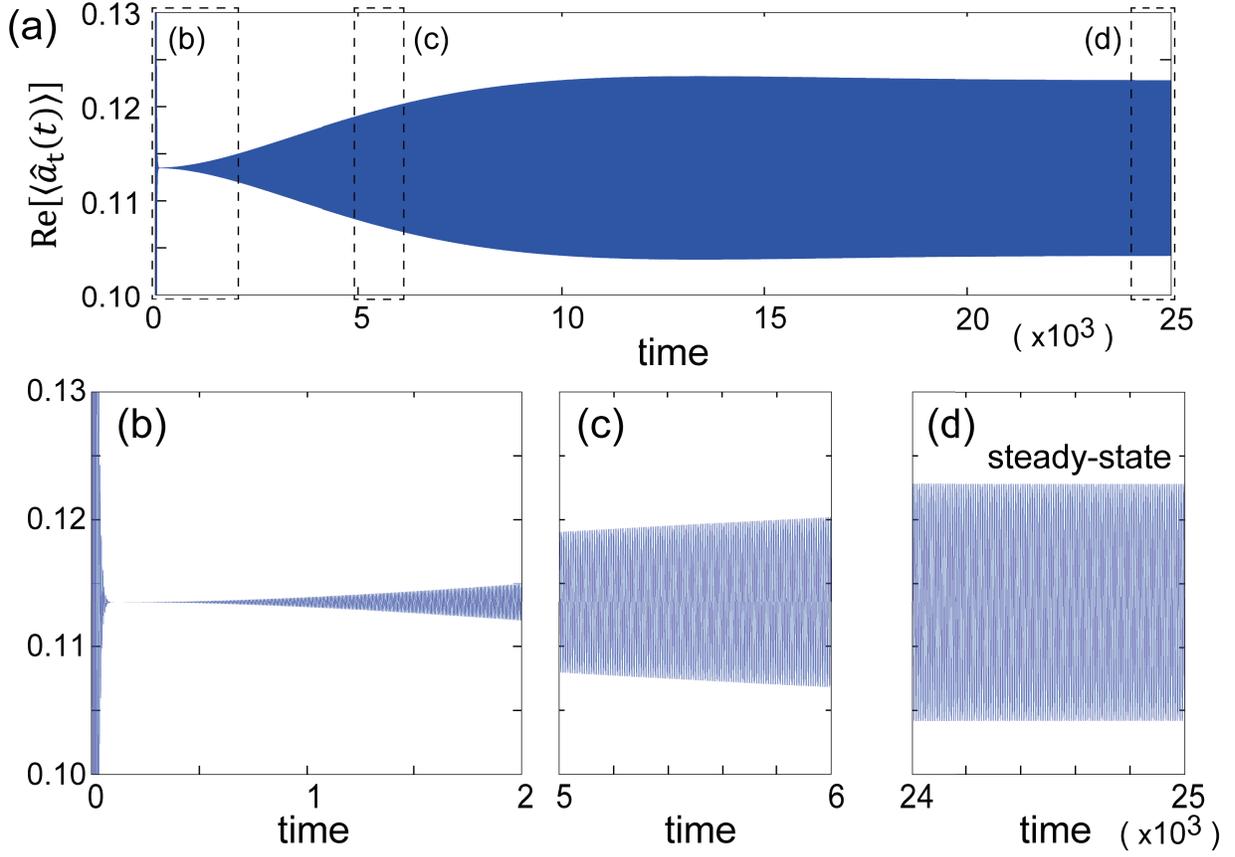}
\caption{
(a) Numerically calculated results of the time evolution of a real part of $\braket{\hat{a}_{\rm t} (t)}$.
The time is normalized by the mechanical vibration frequency $\Omega_0$.
Panels (b)--(d) are zoomed-in views of (a).
$\braket{\hat{a}_{\rm t} (t)}$ becomes a steady state in (d).
The time evolution of the imaginary part is qualitatively equivalent to the real part, hence it is not plotted.
We used the steady state for the evaluation of the power spectrum $\mathcal{S}_\alpha (\omega)$.
}
\label{fig:evolution}
\end{figure}

The cluster expansion provides
\begin{equation}
\braket{\hat{a}_\alpha^\dagger (0) \hat{a}_\alpha (t)}
= \braket{\hat{a}_\alpha^\dagger (0)} \braket{\hat{a}_\alpha (t)}
+ \Delta \braket{\hat{a}_\alpha^\dagger (0) \hat{a}_\alpha (t)}.
\end{equation}
To evaluate the output power spectrum as
$\mathcal{S}_\alpha (\omega) = \int \braket{\hat{a}_\alpha^\dagger (0) \hat{a}_\alpha (t)} e^{-i\omega t} dt$,
we numerically solve the equations of motion for $\braket{\hat{a}_\alpha (t)}$ and
$\Delta \braket{\hat{a}_\alpha^\dagger (0)\hat{a}_\alpha (t)}$.
The former equation of motion follows Eqs.\ (\ref{eq:Heisenberg21}) and (\ref{eq:Heisenberg22}):
\begin{eqnarray}
\frac{d}{dt} \braket{\hat{a}_{\rm c} (t)}
&=&
\left(i\Omega_0 - \frac{\kappa}{2}\right) \braket{\hat{a}_{\rm c} (t)}
+ i g_0 \left( \braket{\hat{a}_{\rm c} (t) \hat{b}_{\rm c}^\dagger (t)}
    + \braket{\hat{a}_{\rm c} (t) \hat{b}_{\rm c} (t)} \right)
    + \left( E_{{\rm c},0} e^{i\Omega_0 t} + E_{{\rm c},+} \right)
\nonumber \\
&=&
\left(i\Omega_0 - \frac{\kappa}{2}\right) \braket{\hat{a}_{\rm c} (t)}
+ i g_0 \left\{
  \braket{\hat{a}_{\rm c} (t)} \braket{\hat{b}_{\rm c}^\dagger (t)}
  + \Delta \braket{\hat{a}_{\rm c} (t) \hat{b}_{\rm c}^\dagger (t)}
  + \braket{\hat{a}_{\rm c} (t)} \braket{\hat{b}_{\rm c} (t)}
  + \Delta \braket{\hat{a}_{\rm c} (t) \hat{b}_{\rm c} (t)}
\right\}
\nonumber \\
& & \hspace{80mm}
  + \left( E_{{\rm c},0} e^{i\Omega_0 t} + E_{{\rm c},+} \right),
\label{eqa:eom_ac} \\
\frac{d}{dt} \braket{\hat{a}_{\rm t} (t)}
&=&
\left(i\Omega_0 - \frac{\kappa}{2}\right) \braket{\hat{a}_{\rm t} (t)}
+ i g_0 \left( \braket{\hat{a}_{\rm t} (t) \hat{b}_{\rm t}^\dagger (t)}
  + \braket{\hat{a}_{\rm t} (t) \hat{b}_{\rm t} (t)} \right)
  + E_{{\rm t},+}
\nonumber \\
&=&
\left(i\Omega_0 - \frac{\kappa}{2}\right) \braket{\hat{a}_{\rm c} (t)}
+ i g_0 \left\{
  \braket{\hat{a}_{\rm t} (t)} \braket{\hat{b}_{\rm t}^\dagger (t)}
  + \Delta \braket{\hat{a}_{\rm t} (t) \hat{b}_{\rm t}^\dagger (t)}
  + \braket{\hat{a}_{\rm t} (t)} \braket{\hat{b}_{\rm t} (t)}
  + \Delta \braket{\hat{a}_{\rm t} (t) \hat{b}_{\rm t} (t)}
\right\}
\nonumber \\
& & \hspace{80mm}
  + E_{{\rm t},+}.
\label{eqa:eom_at}
\end{eqnarray}
For the latter equation of motion, we obtain, e.g.,
\begin{eqnarray}
& &
\frac{d}{dt} \left( \Delta \braket{\hat{a}_{\rm c}^\dagger (0) \hat{a}_{\rm c} (t)} \right)
=
\frac{d}{dt} \left( \braket{\hat{a}_{\rm c}^\dagger (0) \hat{a}_{\rm c} (t)}
- \braket{\hat{a}_{\rm c}^\dagger (0)} \braket{\hat{a}_{\rm c} (t)} \right)
\nonumber \\
& & \hspace{5mm}
= \left(i\Omega_0 - \frac{\kappa}{2}\right) \left\{
\braket{\hat{a}_{\rm c}^\dagger (0) \hat{a}_{\rm c} (t)} - \braket{\hat{a}_{\rm c}^\dagger (0)} \braket{\hat{a}_{\rm c} (t)}
\right\}
\nonumber \\
& & \hspace{10mm}
+ i g_0 \left\{
    \braket{\hat{a}_{\rm c}^\dagger (0)          \hat{a}_{\rm c} (t) \hat{b}_{\rm c}^\dagger (t)}
  - \braket{\hat{a}_{\rm c}^\dagger (0)} \braket{\hat{a}_{\rm c} (t) \hat{b}_{\rm c}^\dagger (t)}
  + \braket{\hat{a}_{\rm c}^\dagger (0)          \hat{a}_{\rm c} (t) \hat{b}_{\rm c} (t)}
  - \braket{\hat{a}_{\rm c}^\dagger (0)} \braket{\hat{a}_{\rm c} (t) \hat{b}_{\rm c} (t)}
\right\}
\nonumber \\   
& & \hspace{5mm}
\approx \left(i\Omega_0 - \frac{\kappa}{2}\right) \Delta \braket{\hat{a}_{\rm c}^\dagger (0) \hat{a}_{\rm c} (t)}
\nonumber \\
& & \hspace{10mm}
+ i g_0 \left\{
    \braket{\hat{a}_{\rm c} (t)}         \Delta \braket{\hat{a}_{\rm c}^\dagger (0) \hat{b}_{\rm c}^\dagger (t)}
  + \braket{\hat{b}_{\rm c}^\dagger (t)} \Delta \braket{\hat{a}_{\rm c}^\dagger (0) \hat{a}_{\rm c} (t)}
  + \braket{\hat{a}_{\rm c} (t)}         \Delta \braket{\hat{a}_{\rm c}^\dagger (0) \hat{b}_{\rm c} (t)}
  + \braket{\hat{b}_{\rm c} (t)}         \Delta \braket{\hat{a}_{\rm c}^\dagger (0) \hat{a}_{\rm c} (t)}
\right\}.
\end{eqnarray}
Then, we can calculate all the expectation values for a single operator and the quantum correlation of the two operators.
Note that Eqs.\ (\ref{eqa:eom_ac}) and (\ref{eqa:eom_at}) include the quantum correlation of the same time operators.
Therefore, we must consider their equations of motion.
In the evaluation, $\braket{\hat{\mathcal{O}}(0)}$ represents the time-averaged steady-state mean value of the system operator $\hat{\mathcal{O}}(t)$.
In this study, we focus only on the sum of the calculated $\braket{\hat{a}_\alpha^\dagger (0)} \braket{\hat{a}_\alpha (t)}$ and
$\Delta \braket{\hat{a}_\alpha^\dagger (0) \hat{a}_\alpha (t)}$.
These respective terms should be considered when the cooperativity $C$ is large.

\section{Time evolution for steady-state}
\label{appen:evolution}
We demonstrate the time evolution using the Heisenberg equations to determine the steady state.
Figure \ref{fig:evolution}(a) shows the oscillated time evolution of ${\rm Re} [\braket{\hat{a}_{\rm t} (t)}]$ as an example.
At the beginning (Fig.\ \ref{fig:evolution}(b)), the oscillation amplitude of $\braket{\hat{a}_{\rm t} (t)}$ changes rapidly.
After a short period, the amplitude gradually increases in Fig.\ \ref{fig:evolution}(b) and (c).
When we examine a sufficient time evolution, $\braket{\hat{a}_{\rm t} (t)}$ becomes a steady state (Fig.\ \ref{fig:evolution}(d)).
In the main text, Sec.\ \ref{sec:numerical_solution}, the output power spectrum $\mathcal{S}_\alpha (\omega)$ is evaluated from
such a steady-state.


\bibliography{citation}

\end{document}